\documentclass[oneside]{style/cernyrep}
\usepackage[T1]{fontenc}

\usepackage{comment} 
\usepackage{amsmath}  
\usepackage{graphicx}
\usepackage{caption}
\usepackage{chngcntr}
\counterwithin{table}{section}
\counterwithin{figure}{section}
\counterwithin{equation}{section}
\DeclareUnicodeCharacter{2212}{-}
\usepackage{float}
\usepackage{times,lipsum}
\usepackage[margin=1in]{geometry}
\usepackage[onehalfspacing]{setspace}
\usepackage{wrapfig}
\usepackage{xcolor}

\usepackage{subcaption}
\usepackage{subfig}
\usepackage{subfloat}
\usepackage{multirow}
\usepackage{adjustbox}
\usepackage{rotating}
\usepackage[colorlinks=true, linktoc=page, pdftex, linkcolor=blue, citecolor=blue, urlcolor=blue]{hyperref}

\usepackage{siunitx}
\DeclareSIUnit\clight{\text{c}}

\usepackage[backend=biber,sorting=none]{biblatex}

\usepackage{pdfpages}
\usepackage[version=4]{mhchem}
\usepackage{textgreek}
\usepackage{acronym}
\usepackage{balance}
\usepackage{authblk}
\usepackage{fancyhdr}
\usepackage{afterpage}
\usepackage{csvsimple}
\usepackage{tcolorbox}

\usepackage{emptypage}

\usepackage{texnames}
\usepackage{ctable}

\usepackage{blindtext}

\usepackage[toc,page]{appendix}

\usepackage{arydshln} 

\def\CCC{C$^{3}$~}
\def\CCCnospace{C$^{3}$}
\def\CCCfive{C$^{3}$-550}
\def\CCCtwo{C$^{3}$-250}
\def\ee{e^+e^-}

\usepackage[hang,flushmargin]{footmisc}
\fancypagestyle{plain}{}
\newlength{\oddmarginwidth}
\newlength{\evenmarginwidth}
\fancyhfoffset[LO,RE]{\oddmarginwidth}
\fancyhfoffset[LE,RO]{\evenmarginwidth}
\bibliography{bibliography} 

\makeatletter
\providecommand\sf@counterlist{}
\makeatother
\usepackage{tocloft}


\begin{document}

\pagenumbering{roman}
\setcounter{page}{1}


\thispagestyle{empty}
\setlength{\unitlength}{1mm}

\title{ESPPU INPUT: \CCC within the "Linear Collider Vision"\textsuperscript{\textdagger}}

\author{Contact Persons:\\ Emilio Nanni, nanni@slac.stanford.edu\\
Caterina Vernieri, caterina@slac.stanford.edu}
\begin{abstract}
The Linear Collider Vision calls for a Linear Collider Facility with a physics reach from a Higgs Factory to the TeV-scale with $e^+e^{-}$ collisions. One of the technologies under consideration for the accelerator is a cold-copper distributed-coupling linac capable of achieving high gradient. This technology is being pursued by the \CCC collaboration to understand its applicability to future colliders and broader scientific applications. In this input we share the baseline parameters for a \CCC Higgs-factory and the energy reach of up to 3~TeV in the 33~km tunnel foreseen under the Linear Collider Vision. Recent results, near-term plans and future R\&D needs are highlighted.
\end{abstract}


\maketitle
\clearpage
\pagenumbering{arabic}
\setcounter{page}{1}

\setcounter{chapter}{1}
\counterwithout{section}{chapter}
\section{Overview}

\CCC is a concept for a linear collider based on  compact, high-gradient, normal-conducting, distributed-coupling accelerators that are operated at cryogenic temperature.\cite{vernieri2023cool,dasu2022strategy,bai2021c,nanni2023status,nanni2022c} \CCC is designed and optimized with a holistic approach to the main linacs, collider subsystems, and beam dynamics to deliver the required luminosity at the overall lowest cost. \CCC was optimized at 250 and 550~GeV center of mass to allow for a Higgs physics program that could also perform measurements of the Higgs self-coupling. A total facility length of 8~km is sufficient for operation at 250 and 550 GeV center of mass. The initial operation at 250 GeV can be upgraded to 550 GeV simply with the addition of rf power sources to the main linacs. This is possible because the increase in gradient and resulting increased power requirement is balanced with an adjustment to the beam format to maintain constant beam-loading, or fraction of the rf power delivered to the beam, which preserves the rf efficiency of the system. By optimizing the cavity geometry, rf distribution, and operating at liquid nitrogen temperatures where the conductivity of copper is increased significantly, the peak-power requirement from the high-power rf sources is significantly reduced. This allows a high beam loading, nearly 50\%, making for a compact and efficient machine. It is important to emphasize that for a normal conducting machine, the cost driver for the main linacs is the rf power, and through the adoption of distributed-coupling and cold-copper technology, peak rf power requirements are reduced by a factor of 6 compared to NLC.

\begin{table}[h!]
\begin{center}
\begin{tabular}{|c | c | c | c | c | c | c |} 
 \hline
  Scenario &  \CCCtwo & \CCCfive  & \CCCtwo  & \CCCfive & \CCCtwo  & \CCCfive  \\
     &   &   & s.u. &  s.u. &  high lumi &  high lumi \\
   \hline\hline
     Luminosity [x10$^{34}$cm$^{-2}$s$^{-1}$] & 1.3 & 2.4 & 1.3 & 2.4 & 7.6 & 4.8  \\
Gradient [MeV/m] & 70 & 120  & 70 & 120 & 70 & 120 \\
Effective Gradient [MeV/m] & 63 & 108 & 63 & 108 & 63 & 108 \\   %

  Num. Bunches per Train  & 133 & 75 & {266} & {150} & {532} & {300} \\
  Train Rep. Rate [Hz] & 120 & 120 & {60} & {60} & 120 & {60} \\
  Bunch Spacing [ns] & 5.26 &  3.5 & {2.65} &  {1.75} & {2.65} &  {1.75} \\
Single Beam Power [MW]  & 2 & 2.45 & 2 & 2.45 & 8 & 4.9 \\
  Site Power [MW] & $\sim$150 & $\sim$175 & {$\sim$110} & {$\sim$125} & {$\sim$180} & {$\sim$180}  \\ 

   Parameter Set \cite{ntounis2024luminosity} & PS1 & PS2 & PS1 & PS2  & PS2 & PS2  \\

\hline
 \end{tabular}
\end{center}
\caption{Parameter table for \CCC comparing the original, sustainability update (s.u.) and high luminosity (high lumi) options. Bunch charge and crossing angle are constant at 1~nC and 14~mrad. Site length of 8~km with details of layout shown in Fig.~\ref{C3layout}.}
\label{tab:param}
\end{table}

\begin{wrapfigure}{rt}{0.48\textwidth}
   \begin{center}
   \includegraphics[trim= 160 0 160 0,clip,width=0.48\textwidth, angle=180]{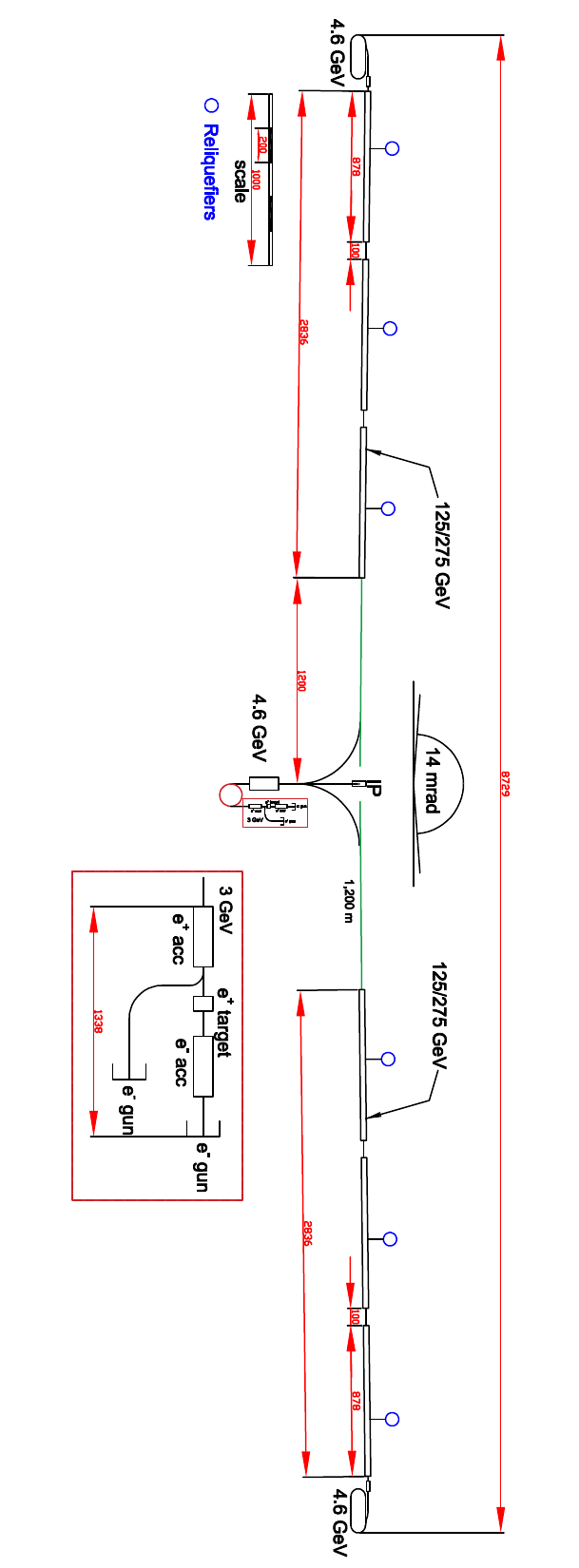}
   \end{center}
   \caption{Major features for the \CCC accelerator complex to scale.}
   \label{C3layout}
\end{wrapfigure}

The current optimized parameters for \CCC are listed in Table \ref{tab:param} with the layout shown in Fig. \ref{C3layout}. The central injector complex, located in the proximity of the interaction point (IP), consists of both the polarized electron source linac and the electron-driven positron source. The positron beam and the electron beam are accelerated in a common linac to 3 GeV. At 3 GeV the positron beam is then injected into a pre-damping ring. The 3 GeV electron beam bypasses the pre-damping ring.  After pre-damping the positrons, the electron and positron beams are accelerated in a common linac to the injection energy of the damping rings at 4.6 GeV. The beams are then separated and transported to the opposite ends of the accelerator complex. At the opposite ends of the accelerator complex, the positron and electron beams are injected into their respective damping rings. 

The damping rings are positioned to allow for a straight path after extraction into the beginning of the linacs to eliminate beam turn-arounds for the damped beams which could degrade the emittance. Upon exiting the damping rings, the beams are transported through the bunch compressors, main linacs and beam delivery systems (BDS) to the IP with a 14 mrad crossing angle. The main linac and BDS are sized for 550 GeV CoM.

\CCC was optimized based on the incredible advances that were made as part of the ILC and CLIC programs in the development of electron and positron sources, injector linacs, damping rings, bunch compressors, beam transport, beam dynamics, beam delivery and detectors. These highly complex and optimized systems serve as the baseline for what is technologically achievable in the linear collider complex and allow for the optimization of the main linac using the \CCC approach with minor modifications to the other sub-systems.

\section{Physics performance}

The beam parameter optimization for \CCC at center-of-mass energies of 250 GeV and 550 GeV was conducted as detailed in \cite{prab_c3}. The primary background from beam-beam interactions that we consider at \CCC is the "incoherent" production, i.e. interactions off of resolved photons rather than fields, of electron-positron pairs. 
The study is based on \texttt{GuineaPig++}  \cite{schulte1999guineapig} simulations of the beam-induced backgrounds and optimize the instantaneous luminosity.

The revised parameter set, referred to as Parameter Set 2 (PS2), achieves a $\sim40\%$ increase in total luminosity compared to the baseline Parameter Set 1 (PS1), while ensuring that bremsstrahlung-induced backgrounds remain at acceptable levels. These are reported in Table~\ref{tab:optimized_params}.

\begin{table}[ht]
\centering
\caption{Baseline (PS1) and optimized (PS2) luminosity parameter sets for C3.}
\label{tab:optimized_params}
\begin{tabular}{@{}lcccc@{}}
\hline
Parameter & \CCCtwo (PS1) & \CCCtwo (PS2) & \CCCfive (PS1) & \CCCfive (PS2) \\
\hline
Horizontal emittance $\epsilon_x^*$ [nm] & 900 & 1000 & 900 & 1000 \\
Vertical emittance $\epsilon_y^*$ [nm] & 20 & 12 & 20 & 12 \\
Vertical waist shift $w_y$ [$\mu$m] & 0 & 80 & 0 & 80 \\
Geometric luminosity [$10^{34}\,\text{cm}^{-2}\text{s}^{-1}$] & 0.75 & 0.92 & 0.93 & 1.14 \\
Total luminosity [$10^{34}\,\text{cm}^{-2}\text{s}^{-1}$] & 1.35 & 1.90 & 1.70 & 2.40 \\
Average Beamstrahlung parameter $\langle \Upsilon \rangle$ & 0.065 & 0.062 & 0.21 & 0.20 \\
\hline
\end{tabular}
\end{table}

A critical part of the optimization involved lowering the vertical emittance and introducing a vertical waist shift ($w_y$) of 80 $\mu$m, enhancing luminosity through better vertical focusing at the IP. 
Additionally, a moderate increase in the horizontal emittance ($\epsilon^*_x$) from 900 nm to 1000 nm effectively controlled beam-beam interaction effects, thus maintaining a manageable luminosity spectrum and limiting detector occupancy from background particles.

\begin{figure}[h!]
   \centering
   \begin{subfigure}[b]{0.49\textwidth}
   \includegraphics[width=\textwidth]{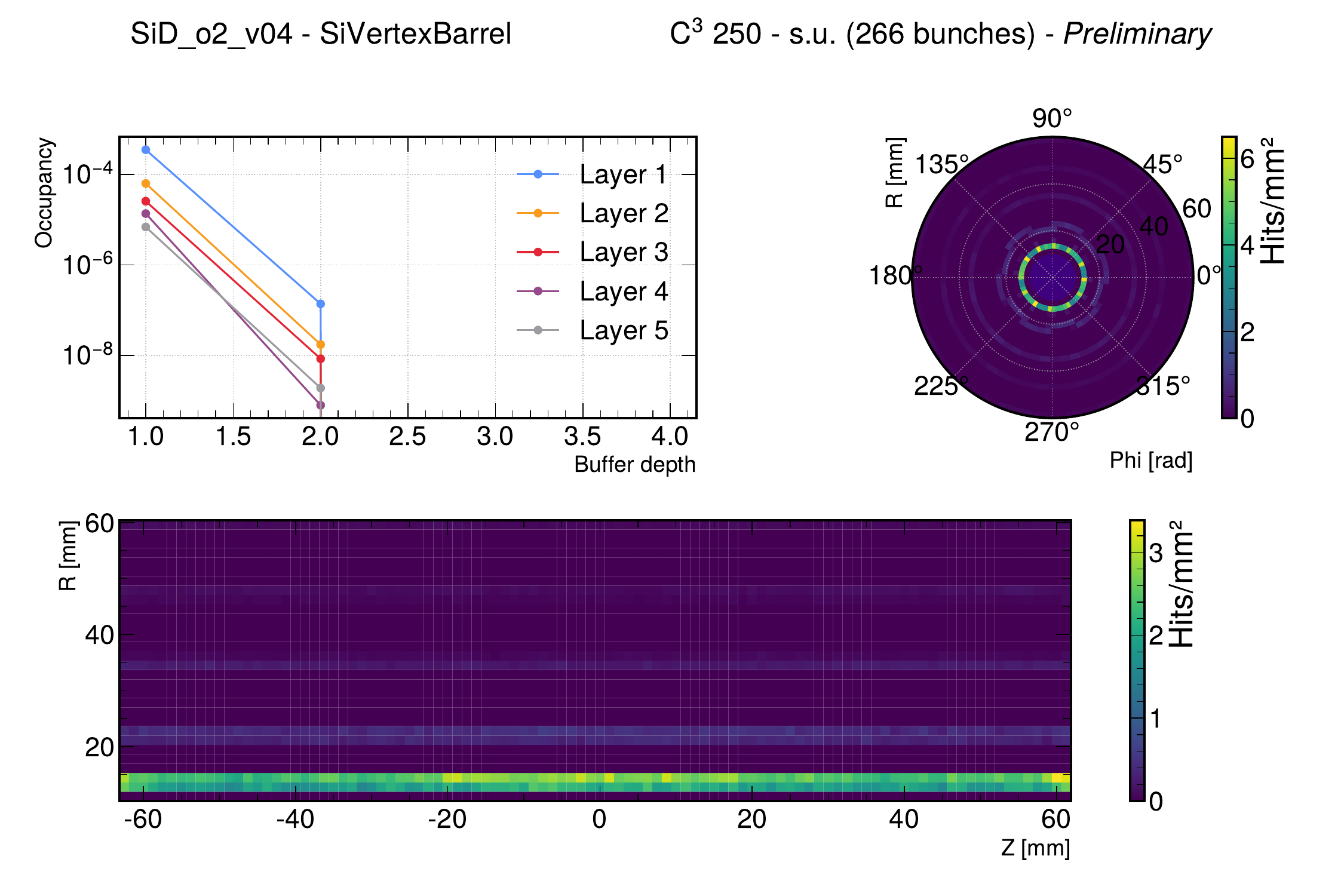}
   \caption{}
   \label{fig:bib250}
   \end{subfigure}
   \begin{subfigure}[b]{0.49\textwidth}
   \includegraphics[width=\textwidth]{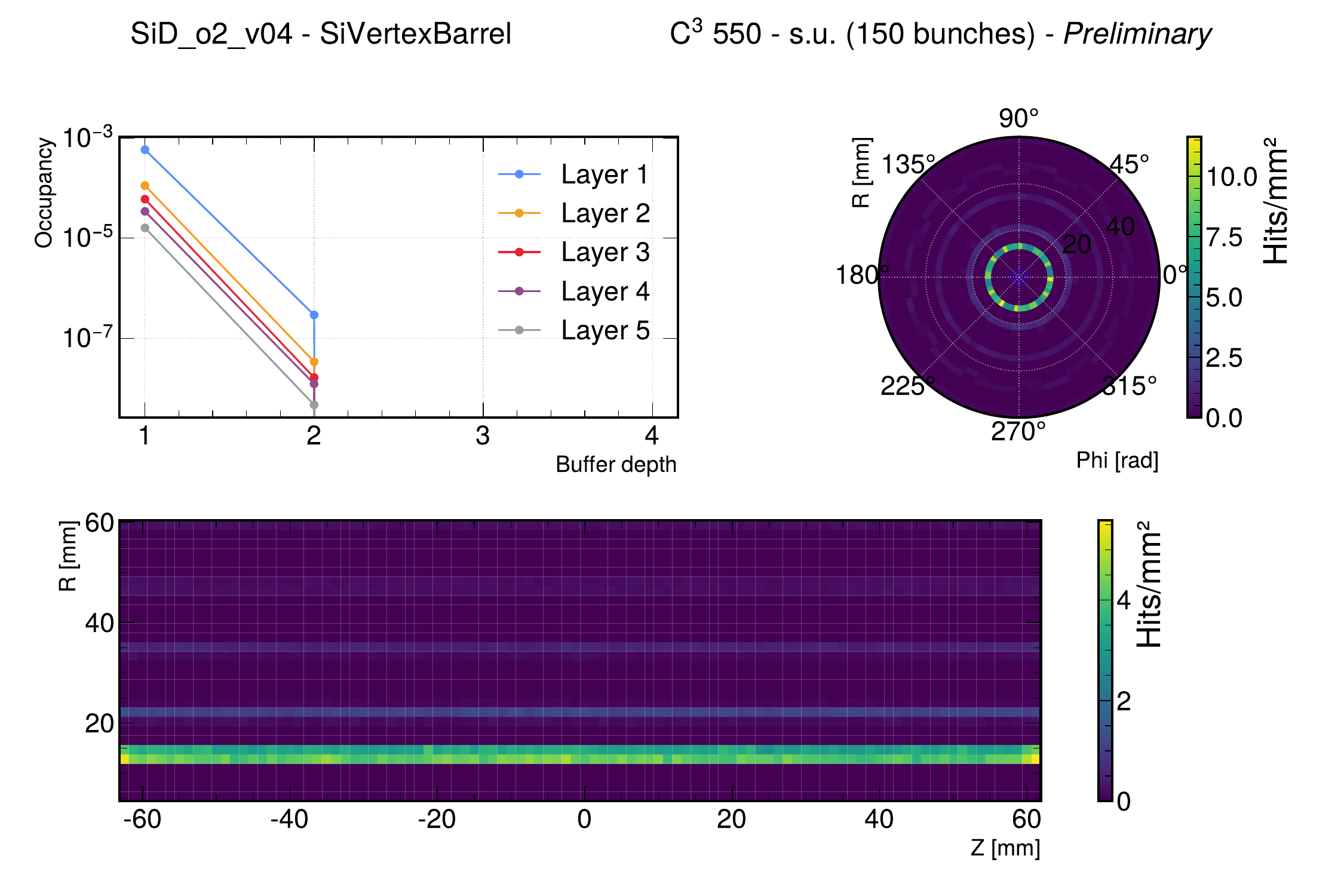}
    \caption{}
   \label{fig:bib550}
   \end{subfigure}
   \caption{Occupancy and hit density distribution for the $e^{+}e^{-}$ pair beam-induced background at the SiD vertex barrel detector for the sustainability set parameters of \CCC operating at (a) 250 GeV and (b) 550 GeV. The occupancy is defined as the number of pixels that have received a number of hits from the background particles equal to or larger than the corresponding buffer depth. For these plots, a pixel pitch of 10 \textmu m is assumed and the hits are integrated over an entire bunch train.}
\end{figure}

The distributions of incoherently produced $\ee$ secondary particles are crucial for estimating beam-induced background (BIB), impacting detector design. For ILC,  because of the lower repetition rate ($5$–$10\ \mathrm{Hz}$), the background particle rates per train are the highest, as discussed in \cite{prab_c3}. Our preliminary studies show that the \CCC beam structure yields BIB levels compatible with ILC-like detectors. In particular in Fig.~\ref{fig:bib250} and ~\ref{fig:bib550} we report the occupancy and hit map distributions for the sustainability upgrade scenarios of \CCCtwo\space and \CCCfive\space respectively, using full detector simulation for the  SiD detector geometry~\cite{Breidenbach:2021sdo}. Background rates are lower at 250 GeV than at 550 GeV due to the increased $\langle \Upsilon \rangle$ parameter at higher energy, with minimal impact from luminosity enhancement scenarios. The energy and momentum of incoherent pairs increases with center-of-mass energy, while their longitudinal boost remains near unity, indicating a strong forward boost. Thus, most background particles are deflected by the detector’s magnetic field, traveling within the beam pipe and avoiding sensitive components, resulting in minimal detector occupancy despite high background particle production~\cite{prab_c3}. 
For \CCCtwo, we expect on average $4.4 \cdot 10^{-3} $ hits/(ns $\cdot$ mm$^2$) $\approx$ 0.023 hits/mm$^2$/BX in the 1st layer of the vertex barrel detector, within the limits set for SiD at ILC. The occupancy in the SiD vertex barrel for the \CCC beam structure at both center-of-mass energies is well within the limits set for ILC.


\section{C$^3$ Performance for the ``Linear Collider Vision"}

The construction of a Linear Collider Facility (LCF) provides us with the opportunity to consider various pathways for future energy upgrades.\cite{abramowicz2025linearcollidervisionfuture} These upgrades may allow for more advanced technologies presently in development to replace the original technological choices if they provide a significant advantage. Ideally, significant portions of the accelerator complex can be re-utilized to leverage the existing common hardware of a LCF and the research and development investment that went into those sub-systems (e.g. injectors, positron source, damping ring, beam delivery).

Indeed, a critical point is that \CCC is compatible with the injector complex and beam delivery system that will be installed in the first phase of a LCF with a superconducting rf (SRF) technology that reaches 250 GeV center of mass in 20~km, as proposed in the Linear Collider Vision (LCV) baseline \cite{LC_sum_25,abramowicz2025linearcollidervisionfuture}. The present design of the damping ring and bunch compressors for \CCC will be presented in Section \ref{sec:DR} and \ref{sec:BC}, along with a comparison to the size and beam format for the ILC damping ring to illustrate this point. Of the 20~km for the LCF, 15~km consist of the main linac for the collider. After completing the initial run of a LCF, we envision the possibility of replacing these 15~km with a cold-copper distributed-coupling linac with an energy reach of up to 2~TeV in the same footprint. 

This linac could operate at any energy below 2~TeV by a combination of bypass lines or by adjusting the gradient and beam format to maintain a constant beam loading and providing a linear scaling in luminosity with the energy and beam power. The parameters for 1 and 2 TeV operation are shown in  Table~\ref{tab:TeVmainlinacparam}, and would be accomplished only with the addition of rf sources, upgrades to the beam delivery system and modifications to the bunch format. No additional civil construction (i.e. additional tunnel length) would be required. Scaling to 3~TeV requires a 33~km facility.

\begin{table}
\begin{center}
\begin{tabular}{|c | c | c | c |  c |} 
 \hline
  Parameter  & Unit & Value  & Value & Value \\
 \hline\hline
Center of Mass Energy   & GeV & 1000 & 2000 & 3000 \\
 \hline\hline
Site Length & km & 20 & 20 & 33 \\
Main Linac Length (per side) & km & 7.5 & 7 & 10.5\\
Accel. Grad. & MeV/m & {75} & {155} & {155}\\
Flat-Top Pulse Length  & ns & {500} &  {195}  &  {195} \\
Cryogenic Load at 77 K & MW & 14 &  20 &  30 \\
Est. AC Power for RF Sources  & MW  & 68 &  65 &  100 \\
Est. Electrical Power for Cryogenic Cooling  & MW  & 81 & 116 & 175  \\
Est. Site Power & MW  & 200 & 230  & 320 \\
RF Pulse Compression & & {N/A} &  {3X} &  {3X} \\
RF Source efficiency (AC line to linac) & $\%$  & {50} & {80} & {80}\\
Luminosity  & x$10^{34}$ cm$^{-2}$s$^{-1}$  & $\sim$4.5 & $\sim$9 & $\sim$14 \\ 
Single Beam Power  & MW & 5 & 9 & 14 \\ 
Bunch Spacing & ns  & 3 &  1.2  &  1.2 \\
\hline
 \end{tabular}
\end{center}
\caption{Main Linac parameters for C$^3$ at 1, 2 and 3~TeV center of mass energy. Bunch charge and repetition rate are 1~nC and 60~Hz for all three cases.}
 \label{tab:TeVmainlinacparam}
\end{table}

\section{Ongoing Component R\&D}

Significant technical progress has been made recently in the development of \CCC technology. In particular, the accelerating structures that can achieve this gradient have been built and demonstrated. Further research and development is needed for the integration of higher order mode damping for the accelerator structures, and cryo-module integration and demonstration that the accelerators and particularly the permanent magnet quadrupoles  meet the alignment and vibration tolerances. A \CCC demonstrator plan was developed to identify the required metrics and experiments that would be needed to advance the technical maturity of the main linac to a level required for a collider design \cite{nanni2023status}. Since this demonstrator plan was published, significant progress has been made on structure design \cite{Schneider:2024kuj,Dhar:2024hgn}, high gradient testing~\cite{liu2025high}, structure vibration measurements \cite{dhar2024vibration}, damping materials \cite{Xu:2024gvb}, alignment system \cite{van2024alignment}, low-level rf \cite{liu2025compact,liu2024direct,liu2024next} and cryo-module design. We continue to work towards the technical goals and timeline set forth in this plan.$^*$

\begin{figure}
   \centering
   \begin{subfigure}[b]{0.45\textwidth}
   \includegraphics[width=\textwidth]{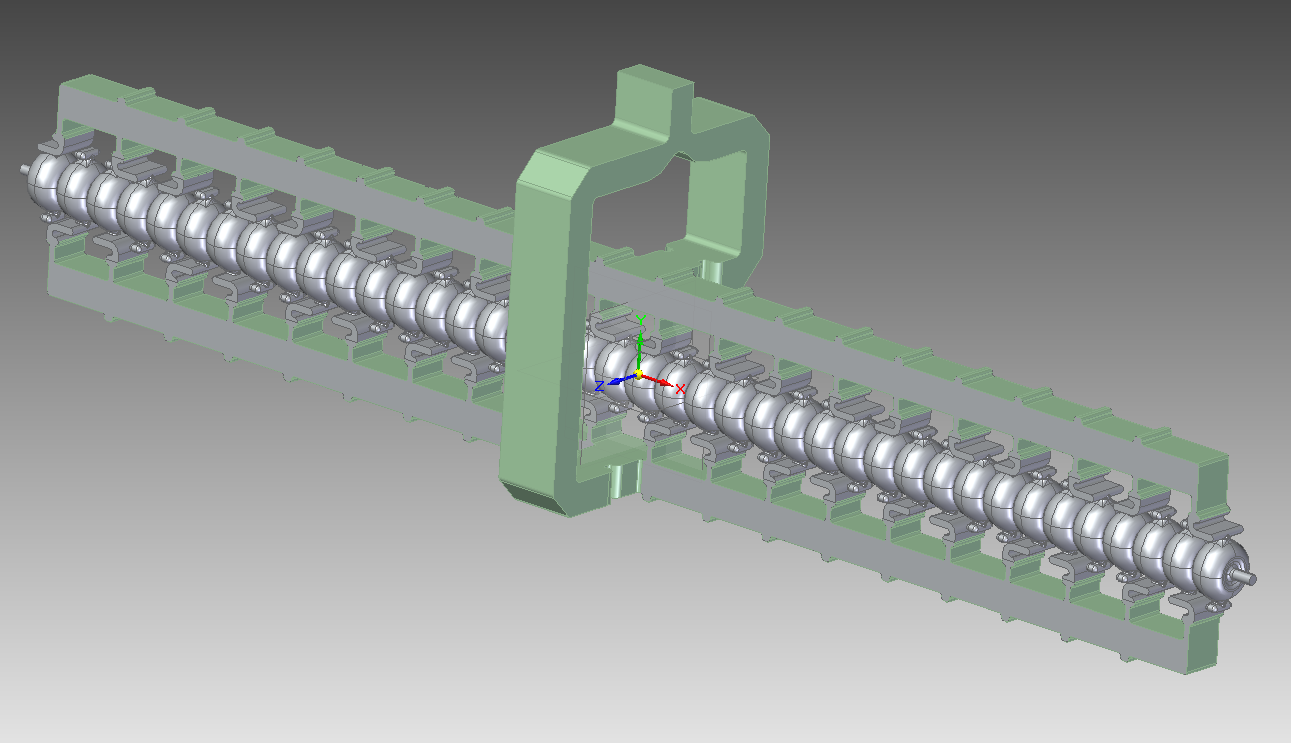}
   \caption{}
   \label{fig:vac_model}
   \end{subfigure}
\begin{subfigure}[b]{0.35\textwidth}
   \includegraphics[width=\textwidth]{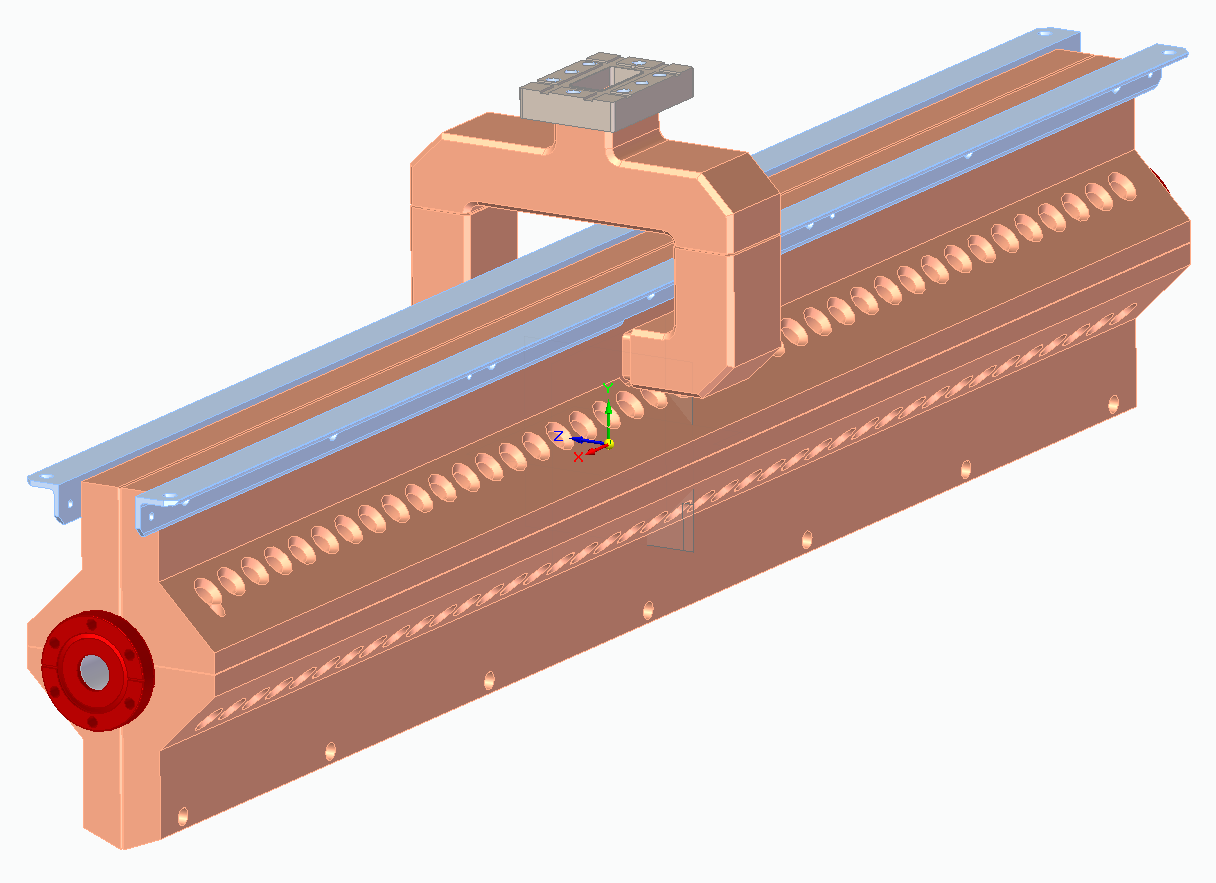}
    \caption{}
   \label{fig:acc_model2}
   \end{subfigure} 
   
   \begin{subfigure}[b]{0.35\textwidth}   \includegraphics[width=\textwidth]{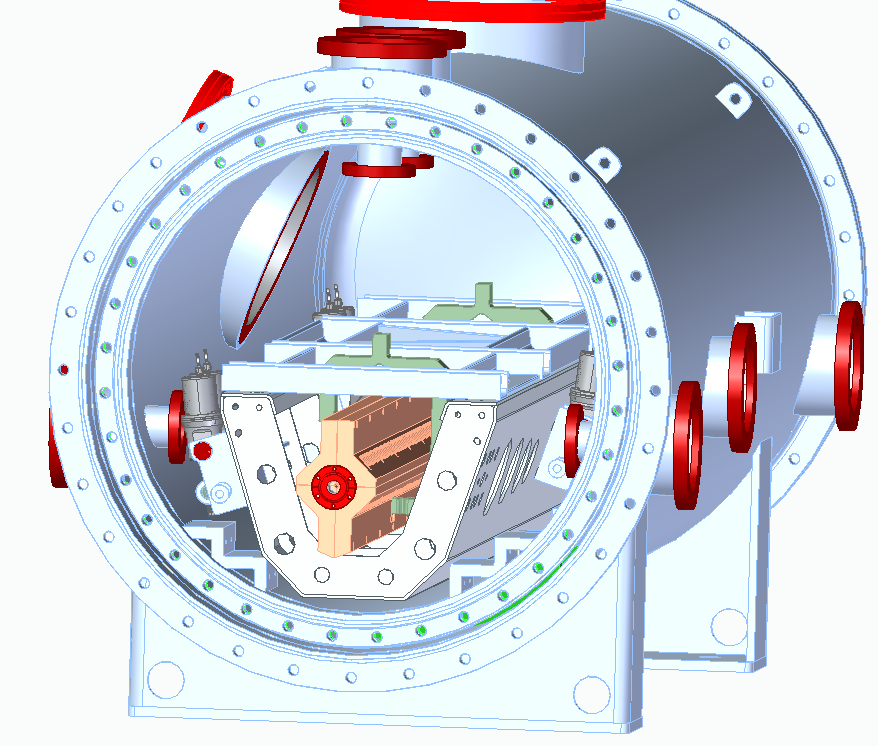}
    \caption{}
   \label{fig:loadedQCM}
   \end{subfigure}
    \begin{subfigure}[b]{0.35\textwidth}
   \includegraphics[width=\textwidth]{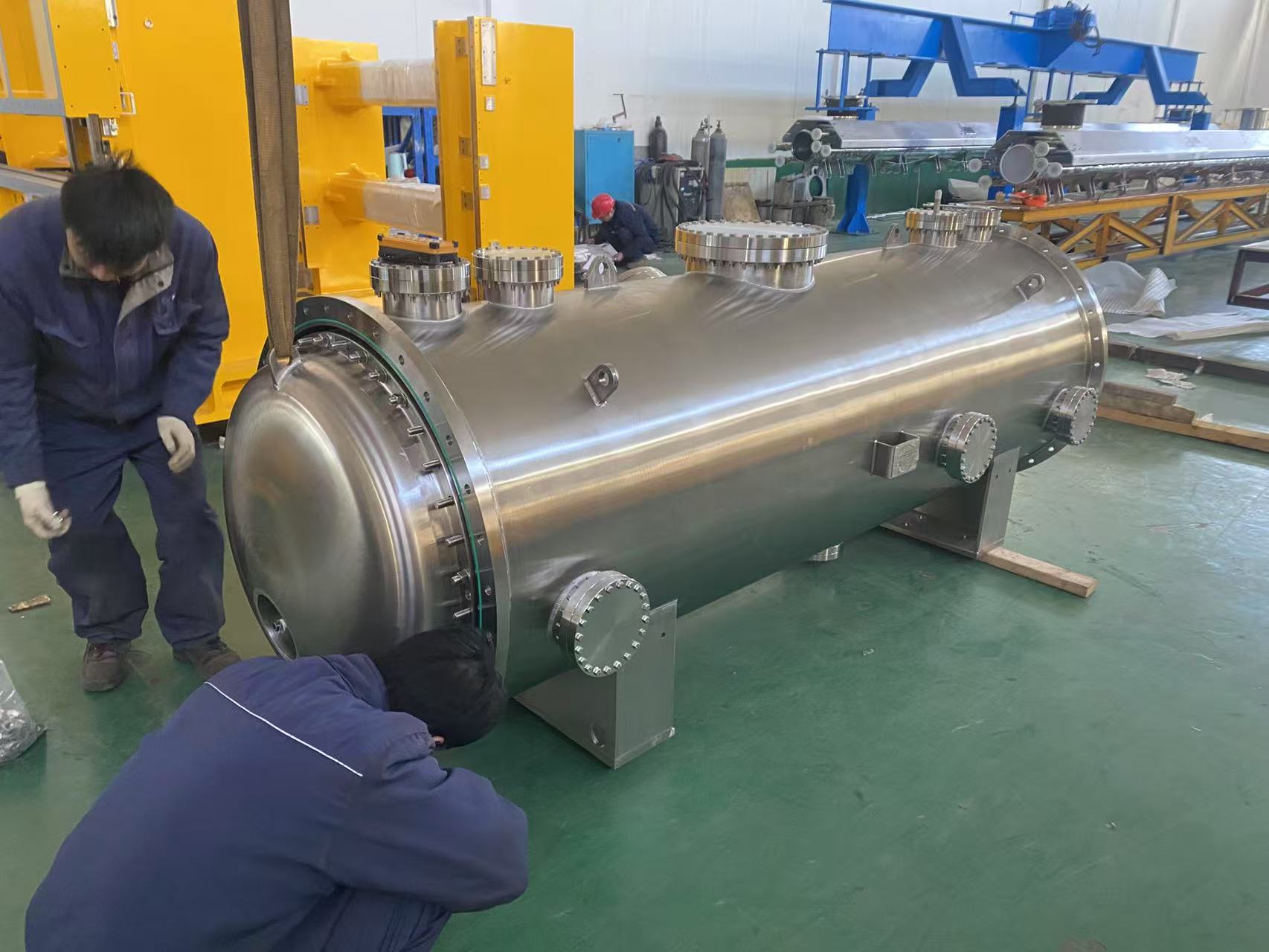}
    \caption{}
   \label{fig:QCM}
   \end{subfigure}
   
   \caption{(a) A 3D rendering of the vacuum space for the large aperture \CCC structure. 
   (b) The solid model of the full structure. For reference the structure length is $\sim$1 m. 
   (c) The QCM with the raft inserted.  The raft is for supporting and loading two accelerating structures and one permanent magnet quadrupole.  (d) The fully completed QCM prior to delivery at SLAC.}
\end{figure}

Ongoing efforts are focused on integration of higher order mode damping, cryo-module design and integration, alignment and vibration measurements. A “quarter cryo-module (QCM)” has been built for tests with radio-frequency (rf) power, beam, and full instrumentation for alignment and vibrations. The QCM will test the raft concept for \CCC\hspace{-0.5em}, shown in Fig~\ref{fig:loadedQCM}. The completed QCM is shown in Fig. \ref{fig:QCM}. The QCM represents an essential step as it contains one of the four repeating elements that make up the full cryo-modules for the main linac. The QCM will also be used to test structures for the injector linac with plans to use the LEA bunker at Argonne. 

On the intermediate timescale, 3-5  full scale cryo-modules, making a few GeV accelerator, are needed to demonstrate technical maturity and get to near production prototypes. Such an R\&D program as proposed at Snowmass is technically achievable before 2030 allowing for \CCC technology to be integrated into plans for an upgrade to the LCF.

\begin{wrapfigure}{rt}{0.5\textwidth}
   \centering
   \begin{subfigure}[b]{0.22\textwidth}
   \includegraphics*[width=\textwidth]{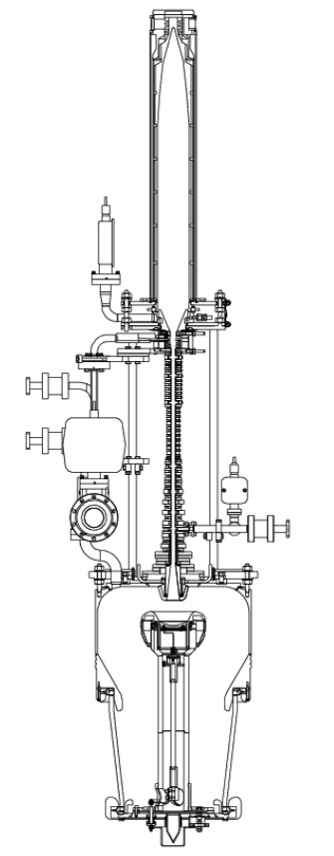}
   \caption{}
   \label{fig:xp3}
   \end{subfigure}
   \begin{subfigure}[b]{0.22\textwidth}
   \includegraphics*[width=\textwidth]{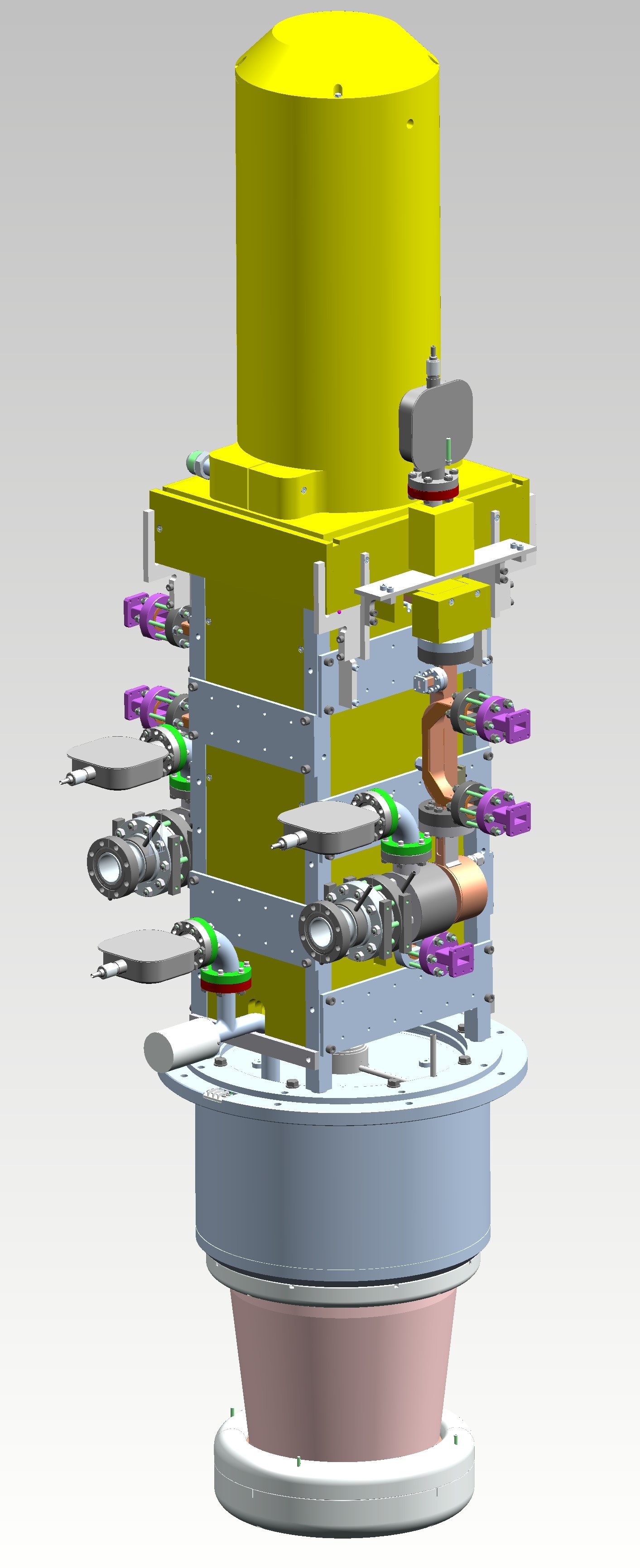}
   \caption{}
   \label{fig:xp3}
   \end{subfigure}
   \caption{(a)Schematic of the XP3 PPM focused 75 MW klystron. (b) 3D model of the new XP4 PPM 75 MW klystron.}
\end{wrapfigure}

The structures for \CCC are also being updated to increase the beam aperture for reduced wakefields and to minimize the effects of vibrations from nucleate boiling of the liquid nitrogen. The manifolds have been rotated to the vertical direction to reduce the horizontal width and avoid trapping of gas. The vacuum space of the cavities, manifold and rf feed is shown in Fig. \ref{fig:vac_model}. The solid model rendering in Fig.  \ref{fig:acc_model2} shows quadrature assembly to allow for the addition of damping slots. The accelerating structure is supported in a raft shown in Fig.  \ref{fig:loadedQCM}. The raft is loaded with two accelerating structures and a permanent magnet quadrupole before being inserted into the QCM, Fig. \ref{fig:loadedQCM}.

Rf source efficiency is also a critical R\&D area for \CCC\hspace{-0.5em}. In particular, reducing the power consumption of the rf source due to the electromagnet for transporting the beam is critical to achieving our power numbers. SLAC is developing an updated periodically focused permanent magnet for transporting beams with sufficient power to produce 80 MW peak power in a C-band rf source.  These periodic permanent magnets (PPMs) are improved designs based on the original PPM XL series for NLC, shown in Fig. \ref{fig:xp3}. This peak power meets the full requirements of the \CCC baseline and LCF scenarios for the rf sources.

\section{Ongoing System R\&D}

To understand the applicability of any technology to the landscape of collider physics  one must investigate the system level performance of the collider complex. Central to achievable luminosity of a linear collider is the preservation of emittance through acceleration and beam delivery. Ongoing \CCC studies aim to increase our technical readiness and fidelity in key performance metrics that will determine the luminosity. Ongoing studies are described for the design of the damping ring in Section \ref{sec:DR}, the bunch compressor in Section \ref{sec:BC}, and the main linac in Section \ref{sec:ML}

\subsection{Damping Ring}
\label{sec:DR}

The goal of the damping ring is to produce a flat beam with as small an emittance as possible subject to the constraint of having a sufficiently fast damping time to match the collider repetition rate, $f$. Generally, multiple damping cycles are required to allow the injected beam to reach its equilibrium emittance. This requirement along with the collider bunch structure which informs the choice of ring circumference, sets roughly the beam energy:
\begin{equation}
\tau_y=\frac{3C}{r_ec\gamma^3I_2},
\end{equation}
where $C$ is the ring circumference, $\gamma$ is the relativistic factor, $r_e$ is the classical electron radius and $I_2$ is the second radiation integral.

\begin{figure}
   \centering
   \begin{subfigure}[b]{0.45\textwidth}
   \includegraphics*[width=\textwidth]{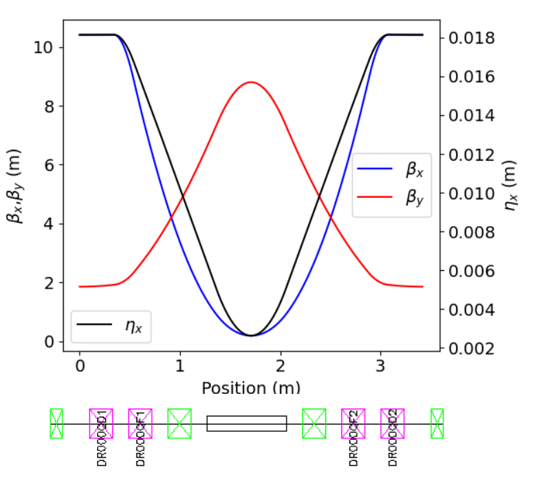}
   \caption{}
   \label{TME}
   \end{subfigure}
   \begin{subfigure}[b]{0.45\textwidth}
   \includegraphics*[width=\textwidth]{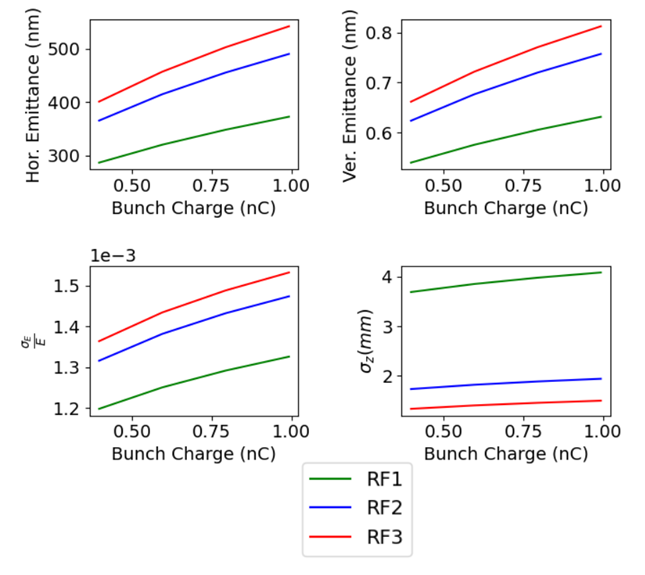}
   \caption{}
   \label{IBS}
   \label{fig:xp3}
   \end{subfigure}
   \caption{(a) Beta functions and dispersion for a TME cell. (b) Horizontal, vertical normalized emittances, RMS relative energy spread and bunch length as a function of charge due to IBS. Each color refers to a different rf voltage in the ring.}
\end{figure}

For the C$^3$ collider the main electron and positron damping rings will have a racetrack layout. The arcs are comprised of Theoretical Minimum Emittance (TME) cells while the straight sections are used for injection and extraction and to provide space for damping wigglers. An analytic study varying the major parameters (beam energy, number of stored bunches, wiggler strengths and lengths) was performed and a choice of storing two bunch trains was made as a trade-off between size of the machine and achievable emittance resulting in an energy of 4.6 GeV. Each bunch train is stored for approximately 7.9 damping times. While this number is relatively large, lowering it by reducing the beam energy results in a larger final emittance when Intra-Beam Scattering (IBS) is accounted.

The zero-charge horizontal emittance is largely determined by the number of TME cells in the ring:
\begin{equation}
\gamma\epsilon_x=\frac{C_q\gamma^3\theta^3}{12\sqrt{15}},
\end{equation}
where $\gamma\epsilon_x$ is the normalized horizontal emittance, $C_q$ is a numerical constant and $\theta$ is the bend angle per TME cell. The above implies 248 cells are required for a baseline zero-charge normalized emittance of 65 nm. However, at optimal emittance TME cells require tight focusing which introduces a large chromaticity in both planes. We therefore moved off of the optimum settings until the normalized chromaticity in each plane was $<$3 resulting in a zero-charge emittance of 170 nm. For a cell length of 3.4 m the total arc length is 843 m. Each straight section is 172 meters long with the majority of that length used for wiggler magnets, but additional space is also included for beam extraction/injection, rf cavities to replenish energy loss from synchrotron radiation and optics for matching the straight sections and arcs and zeroing dispersion. 

In the current design the ring circumference is 1190 m which fits 2.8 bunch trains. It may be possible to reduce the ring circumference by making the TME cells more compact but for an initial study a conservative estimate of 3.4 m was used. Note that reducing the number of cells would increase the emittance which is not preferable. While another possibility may be to increase the circumference to fit 3 bunch trains.

With the major design parameters selected a model of the ring was produced in BMAD. A code available in the BMAD ecosystem to model IBS based on the Completely Integrated Modified Piwinski (CIMP) formalism was used (see Fig. \ref{IBS}) Final normalized emittances range from 300-500 nm in the horizontal plane depending on rf voltage and bunch charge, while the vertical emittance remains below 1 nm for all rf settings and bunch charges considered. It is important to note that the vertical emittance value will likely be dominated by magnet tolerances and not IBS, which is an ongoing study. These studies already indicate that the baseline \CCC tolerances are met with tolerances that match present fourth generation light sources, giving high confidence in not negative impact on baseline luminosity.

\subsection{Bunch Compressors}
\label{sec:BC}

Particle bunches leaving the damping ring (DR) have been damped strongly in the transverse plane but are too long to be accepted into the main linac where they will be boosted to collision energy. Therefore, a bunch compression scheme between the damping ring and main linac is required. Because the beam energy leaving the DR is 4.6 GeV, ballistic compression is not an option. Instead, the compressor consist of two parts (i) an rf linac which imparts an energy-position correlation (energy chirp) on the bunch and a magnetic chicane which has a path length that depends strongly on a particle's energy. 

Leaving the damping ring, the particle bunch is uncorrelated in position with a bunch length of the order of a few mm and a relative energy spread of approximately 0.1$\%$ (depending on bunch charge and rf setting in the DR). The compression goal is 100 $\mu$m. In the absence of beam intensity effects (in particular Coherent Synchrotron Radiation (CSR)) the longitudinal emittance is conserved which implies an energy spread on the order of 1 $\%$ is expected after compression.

Analytic formulas derived from linear transfer matrices of an rf cavity, bend magnet, and drift were used to determine the required $M_{65}$ of the rf system and $M_{56}$ of the chicane for the case of operating at the rf zero-crossing. To compensate for a non-negligible  $T_{566}$ from the chicane the beam is moved slightly off the zero-crossing to take advantage of the curvature present in the rf field. This results in a slight deceleration of the beam from 4.6 GeV to about 3.9 GeV. 

To introduce the energy chirp, an rf frequency of 2.85 GHz was chosen since it is a sub-harmonic of the main linac. As a reasonable starting point a 15-cell cavity with gradient of approximately 9 MV/m was assumed. To achieve the required energy chirp for a 100 $\mu$m final bunch length, 80 total rf cavities are needed resulting in an approximately 130 m long rf section. Dipole magnets were chosen to be 2 meters in length separated by 4 meter drifts. Typical strengths are  $0.5$ T resulting in a bending angle of 4.5 degrees per dipole. For focusing, a FODO cell comprising the length of 8 rf cavities is used. The chicane does not contain any quadrupoles. 4 quadrupoles on each side of the chicane were tuned to keep the beta functions under 30 m with relatively weak divergence through it.

\begin{wrapfigure}{rt}{0.55\textwidth}
   \centering
   \includegraphics*[width=0.5\columnwidth]{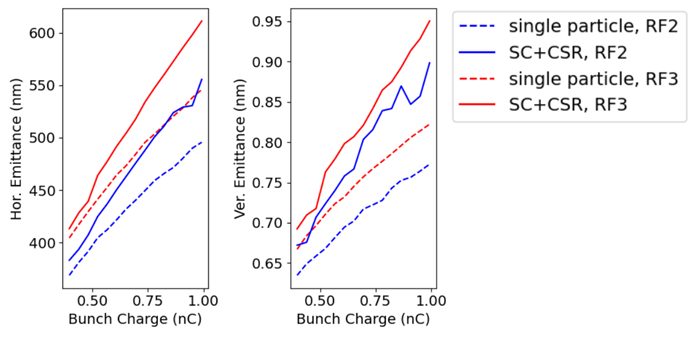}
   \caption{The horizontal and vertical normalized emittances at the end of the bunch compressor. Red and Blue are for two different rf settings in the DR while solid lines include a 1D-CSR calculation while dashed lines do not.}
   \label{CSR}
\end{wrapfigure}
The effect of CSR was modeled using BMAD's built-in 1D CSR calculation. Tracking was done with 10K particles and 100 longitudinal bins (see Fig. \ref{CSR}). We are currently looking into 3D CSR effects. However, we note that due to the small transverse emittance and well controlled beta functions in the chicane, the common rule-of-thumb for when 3D effects become non-negligible $\sigma_{x,y} < \rho^{1/3}\sigma_{Z}^{2/3}$ suggests that the 1D model should be accurate.

\subsection{Main Linac}
\label{sec:ML}

Significant effort has been put into modeling the transport of the electron beam in the \CCC main linac. This study has focusing on understanding a beam optics configuration that transports the electron beam with less than 10\% emittance growth, is tunable with dispersion free steering, and is able to tolerate both short-range and long-range wakefield effects \cite{tan2024multi}. These studies focused on a beam injected into the main linac with the nominal emittance parameters. However, it is also worth exploring the possibility of pursuing lower emittance injection into the main linac informed by the performance of the damping rings and the bunch compressor. Initial studies have been performed assuming a normalized vertical emittance of 1~nm which is the most aggressive expected emittance from the damping ring and bunch compressor, Fig. \ref{CSR}.

\begin{wrapfigure}{rt}{0.5\textwidth}
   \centering
   \includegraphics[trim= 280 90 280 150,clip,width=0.5\columnwidth]{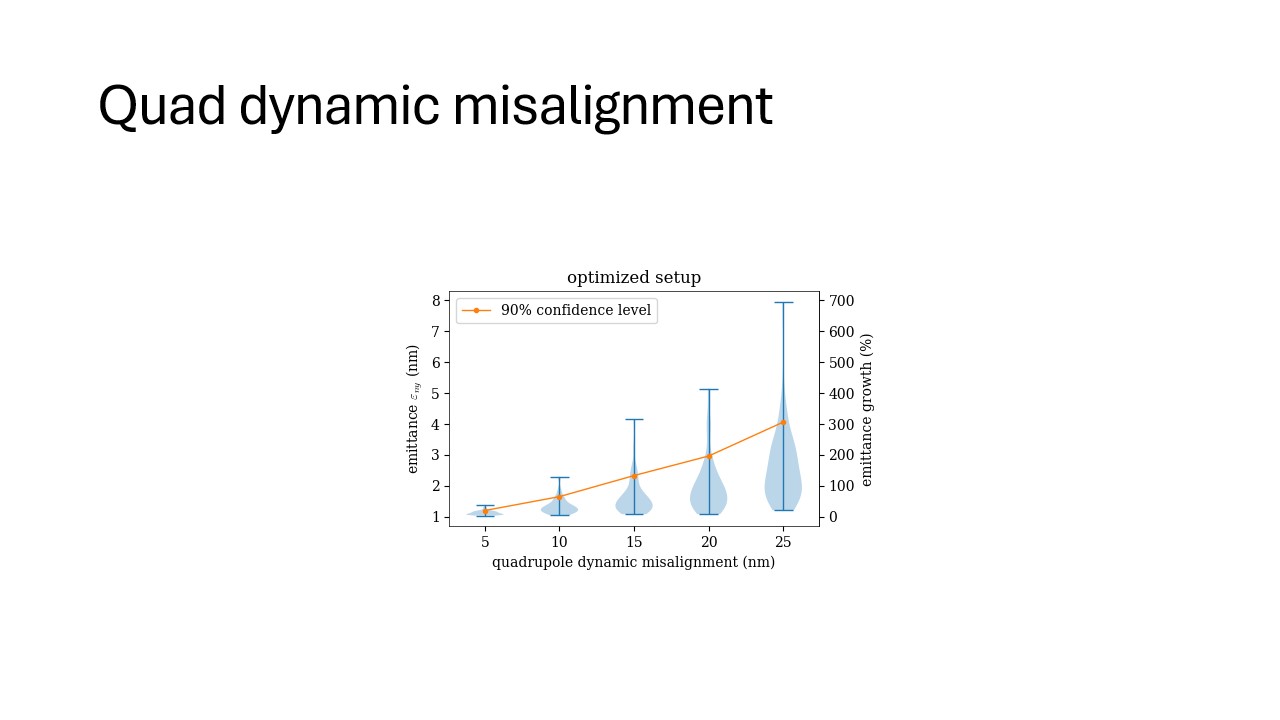}
   \caption{Emittance growth due to quadrupole dynamic misalignment.}
   \label{fig:ml}
\end{wrapfigure}

The results of this study for the main linac, Fig. \ref{fig:ml}, show that excluding IBS the relative vibration tolerances for injection, the structures and the quadrupoles are unchanged. Future studies will include matching the full 6D phase space of the beam from the damping rings and bunch compressor at injection, including expected variations in beam parameters due to the injector complex, full single- and multi-bunch beam dynamics, as well as IBS.

\subsection{HOM studies and HOM suppression}

\begin{wrapfigure}{rt}{0.45\textwidth}
   \centering
   \includegraphics*[width=0.45\columnwidth]{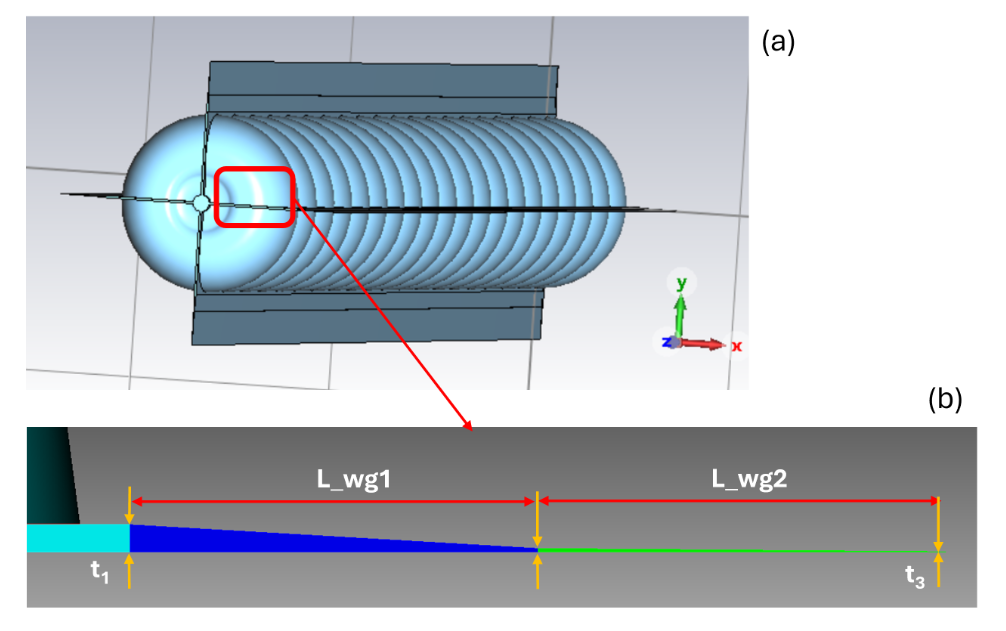}
 \caption{(a) Vacuum space showing cavities and damping slots. (b) Zoom-in view of damping slot showing profile.}
 \label{fig:damp}
\end{wrapfigure}

Recent studies for the main linac have focused on understanding the requirements for damping and detuning to maintain the beam emittance during acceleration \cite{kim2022design,tan2024multi,Xu:2024gvb,kim2023study}. The approach that is being considered for damping of structures is the use of parallel-plate waveguide slots cut in quadrature for the structure, Fig. \ref{fig:damp}. The shape of the damping slots is specially designed to provide effective damping of all HOMs other than the TM$_{0N}$ modes. The slots are coated with nickle-chromium (NiCr) which is a lossy metal whose conductivity does not increase as temperature decreases. The coating and brazing process has been tested on copper samples with measurements at cryogenic temperatures. A two cell structure to test the damping slot approach has been designed and fabricated. \cite{LANLDAMP} The structure will be tested a LANL. 

\section{Synergies}

Cold copper technology is widely applicable to the advancement of particle accelerators for both High Energy Physics, as well as the broader scientific, medical and industrial use of particle accelerators. Cold copper and distributed coupling linacs are extremely efficient and cost effective technologies for reaching high gradients. We are exploring the possibility of utilizing this technology to improve the FCC-ee injector \cite{nanni2024_FCC} and the ILC electron-driven positron scheme. As a main linac technology, cold-copper is also being considered by the HALHF collaboration for the positron linac \cite{adli2025halhfhybridasymmetriclinear}. Cold copper technology is also very appealing for increasing the gradient of the muon cooling channel and is under investigation for the Muon Collider \cite{berg2024engineering}. 

Beyond high energy physics, many applications of cold copper technology are being explored \cite{CCATA25}. In the context of the development of cold-copper for \CCCnospace, injector linacs for synchrotron light sources, linacs for compact FELs \cite{rosenzweig2020ultra,rosenzweig2024high} or high-gradient photoinjectors \cite{lawler24high} are particularly appealing. Ongoing studies are exploring this technology for application of the synchrotrons at Cornell \cite{maxson} and ESRF \cite{simone2025}.

\section{Sustainability \& Environmental Considerations}

Due the the presence of liquid nitrogen in the tunnel, the presence of humans during the cold state of the machine will require proper safety measures. These include adequate ventilation, oxygen deficiency monitors, etc. Requirements will need to be considered concerning relevant regulations.

The overall design of \CCC is being optimized to minimize the cost and impact on the environment. The small physics footprint of \CCC makes it possible to consider time and cost savings, such as surface site construction, that also reduce impacts from waste material from excavation (e.g. cut and cover) and avoid vertical shafts. The presently estimated environmental impact of \CCC is described in \cite{breidenbach2023sustainability}.

\newpage

{\textsuperscript{\textdagger}Authors: Matthew B. Andorf$^{2}$, Mei Bai$^{5}$, Pushpalatha Bhat$^{3}$, Valery Borzenets$^{5}$, Martin Breidenbach$^{5}$,  Sridhara Dasu$^{7}$, Ankur Dhar$^{5}$,  Tristan du Pree$^{6}$, Lindsey Gray$^{3}$, Spencer Gessner$^{5}$, Ryan Herbst$^{5}$, Andrew Haase$^{5}$, Erik Jongewaard$^{5}$, Dongsung Kim$^{4}$, Anoop Nagesh Koushik$^{6}$, Anatoly K. Krasnykh$^{5}$, Zenghai Li$^{5}$, Chao Liu$^{5}$, Jared Maxson$^{2}$, Julian Merrick$^{5}$, Sophia L. Morton$^{5}$, Emilio A. Nanni$^{5}$,   Alireza Nassiri$^{1}$, Cho-Kuen Ng$^{5}$, Dimitrios Ntounis$^{5}$,  Mohamed A. K. Othman$^{5}$, Marco Oriunno$^{5}$, Dennis Palmer$^{5}$,   Michael E. Peskin$^{5}$, Caterina Vernieri$^{5}$, Muhammad Shumail$^{5}$, Evgenya Simakov$^{4}$, Emma Snively$^{5}$, Ann Sy$^{5}$, Harry van der Graaf$^{6}$, Brandon Weatherford$^{5}$ and Haoran Xu$^{4}$}

\noindent
$^{1}${Argonne National Laboratory}\\
$^{2}${Cornell University}\\
$^{3}${Fermi National Accelerator Laboratory} \\
$^{4}${Los Alamos National Laboratory}\\
$^{5}${SLAC National Accelerator Laboratory, Stanford University}\\
$^{6}${Nikhef}\\
$^{7}${University of Wisconsin, Madison}\\

$^*$ \CCC R\&D Timeline:\begin{figure}[h]
   \centering
   \includegraphics[width=\textwidth]{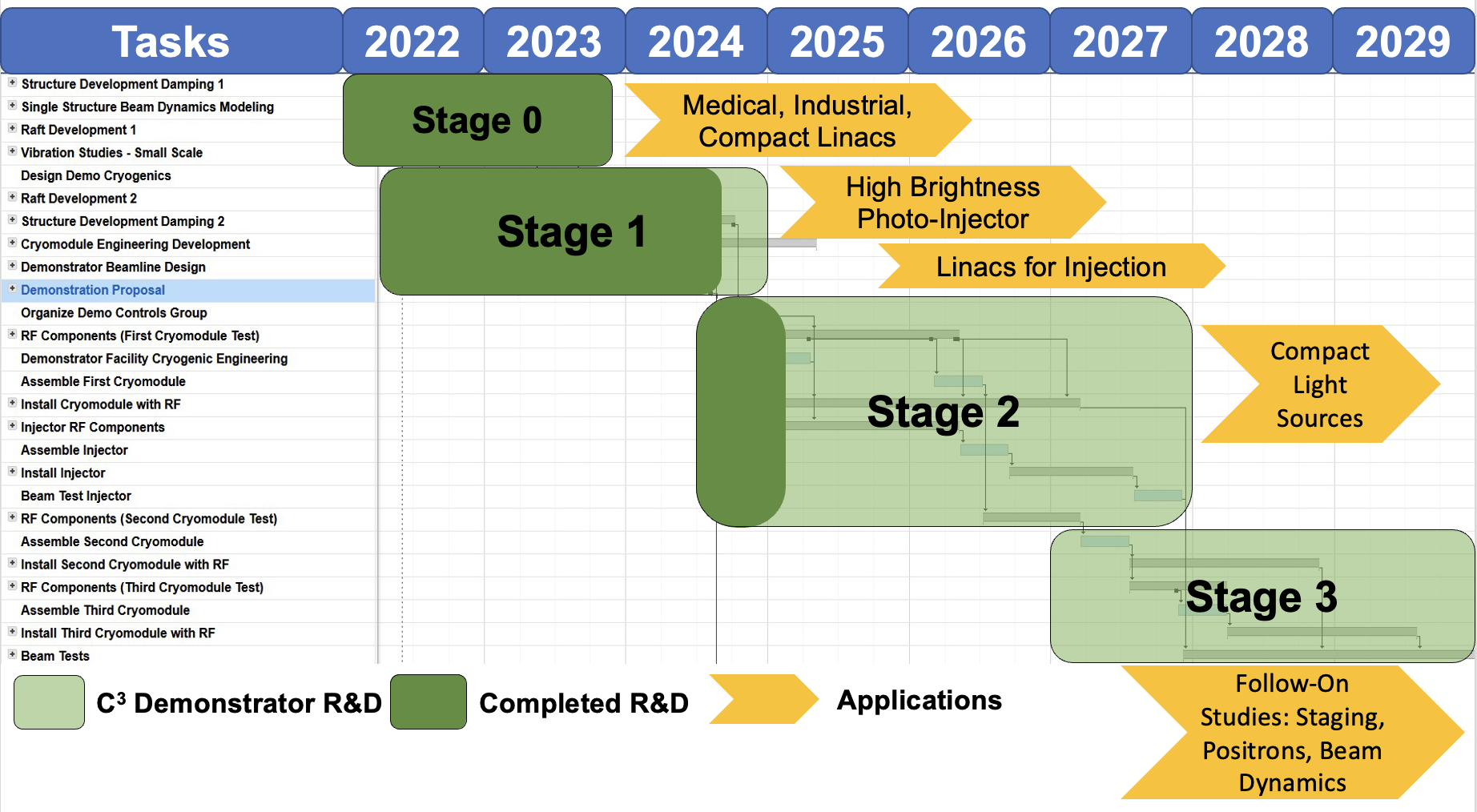}
 
\end{figure}

\printbibliography

\end{document}